\documentclass[12pt]{article}
\usepackage[ascii]{inputenc}
\usepackage[T1]{fontenc}
\usepackage{amsmath}
\usepackage{comment}
\usepackage{icomma}
\usepackage{setspace}
\usepackage{graphicx}
\usepackage{booktabs}
\usepackage{siunitx} 
\date{\today}
\usepackage{float}

\usepackage{tikz}
\usepackage{pgfplots}
\pgfplotsset{compat=1.18}
\usepgflibrary{fpu}
\usepgfplotslibrary{statistics}
\pgfmathdeclarefunction{gamma}{1}{\pgfmathparse{exp(gammaln(#1))}}
\usetikzlibrary{math}

\pgfplotsset{compat=1.18}
\pgfmathdeclarefunction{erlang}{3}{%
  \pgfmathparse{(#2>=0)*((#2^(#1-1))*exp(-#2/#3)/((#3^#1)*gamma(#1)))}%
}

\usepackage{amsfonts} % For \mathbb
\usepackage{geometry} % Optional: for page margins
\newcommand{\E}{\mathbb{E}}
\newcommand{\logit}{\operatorname{logit}} % Define logit operator

\usepackage{lmodern} % monospace font
\usepackage[scale=0.89]{tgheros} % Helvetica is too big
\usepackage{Baskervaldx} % tosf in text, tlf in math;; use [osf] to get old time numbers -- better without old time numbers for %tech document
\usepackage[baskervaldx,cmintegrals,bigdelims,vvarbb]{newtxmath} % math italic letters from Baskervaldx
\usepackage[cal=boondoxo]{mathalfa} % mathcal from STIX, unslanted a bit

\usepackage{hyperref}
\hypersetup{
    colorlinks=true,
    citecolor=blue,
    linkcolor=red,
    filecolor=magenta,      
    urlcolor=blue,
}

\newcommand{\mybib}{\vspace{0.05 in}\setlength{\hangindent}{0.35in}\noindent\samepage }

\title{\textbf{Will Neural Scaling Laws Activate Jevons' Paradox in AI Labor Markets? A Time-Varying Elasticity of Substitution (VES) Analysis}}

\author{{ Rajesh P. Narayanan}\thanks{Louisiana State University, E.\ J.\ Ourso College of Business (\href{mailto:rnarayan@lsu.edu}{rnarayan@lsu.edu}) } \\ \\
\and
{\ R.\ Kelley Pace }\thanks{Louisiana State University, E.\ J.\ Ourso College of Business (\href{mailto:kpace@lsu.edu}{kpace@lsu.edu}) }}

\date{\today}

\begin{document}
\maketitle
\thispagestyle{empty}
\clearpage

\begin{abstract}
\large
We develop a formal economic framework to analyze whether neural scaling laws in artificial intelligence will activate Jevons' Paradox in labor markets, potentially leading to increased AI adoption and human labor substitution. By using a time-varying elasticity of substitution (VES) approach, we establish analytical conditions under which AI systems transition from complementing to substituting for human labor. Our model formalizes four interconnected mechanisms: (1) exponential growth in computational capacity ($C(t) = C(0) \cdot e^{g \cdot t}$); (2) logarithmic scaling of AI capabilities with computation ($\sigma(t) = \delta \cdot \ln(C(t)/C(0))$); (3) declining AI prices ($p_A(t) = p_A(0) \cdot e^{-d \cdot t}$); and (4) a resulting compound effect parameter ($\phi = \delta \cdot g$) that governs market transformation dynamics. We identify five distinct phases of AI market penetration, demonstrating that complete market transformation requires the elasticity of substitution to exceed unity ($\sigma > 1$), with the timing determined primarily by the compound parameter $\phi$ rather than price competition alone. These findings provide an analytical framing for evaluating industry claims about AI substitution effects, especially on the role of quality versus price  in the technological transition.
\end{abstract}

\clearpage
\large
\setstretch{1.2}

\section{Introduction}\label{introduction}

Many industry leaders appeal to Jevons' Paradox (1866) -- the principle that efficiency gains that lower the price of a resource can increase its overall consumption -- when discussing AI progress. Furthermore, industry leaders state that AI scaling laws that deliver quality improvements with increased computation (Kaplan et al., 2020;  Hoffmann et al., 2022) will drive usage.\footnote{For example, Sam Altman (2025), in his blog entitled ``Three Observations,'' stated that ``1. The intelligence of an AI model roughly equals the log of the resources used to train and run it. 2. The cost to use a given level of AI falls about 10X every 12 months, and lower prices lead to much more use. 3. The socioeconomic value of linearly increasing intelligence is super-exponential in nature.'' The first point restates the AI scaling law while the second point is consistent with Jevons' Paradox. Satya Nadella (2025) stated that ``Jevons paradox strikes again! As AI gets more efficient and accessible, we will see its use skyrocket, turning it into a commodity we just can't get enough of.''} Decreasing marginal costs of effective computation (which includes software as well as hardware advances), coupled with competition linking these lower costs to lower AI prices, are seen as key drivers.\footnote{Ho et al. (2024) report that the ``... compute required to reach a set performance threshold has halved approximately every 8 months ...  .'' } Is Jevons' Paradox mere industry jargon, or  does it provide economically meaningful insights relevant to AI progress?

 We develop a partial equilibrium framework that incorporates these statements to clarify the underlying logic of AI adoption. Specifically, we examine market transitions to AI-produced goods using an elasticity of substitution framework. We demonstrate conditions under which complete market transformation occurs, highlighting the role of both price elasticities and substitution elasticities.

 Jevons' Paradox  occurs when increased efficiency in a resource leads to greater total demand due to both \emph{intensive} margins (existing users consuming more) and \emph{extensive} margins (new users or applications entering the market, a process that may take time). For example, the extensive margin of uses from electricity has dwarfed the initial uses of electricity for lighting, but many uses took many years to emerge. In elasticity terms, this implies a long-term price elasticity of demand less than $-1$ (i.e., highly elastic, $E < -1$), where the percentage increase in quantity demanded exceeds the percentage decrease in price. 
 
However, industry statements seem to point to obtaining a larger expenditure share of the entire market from the consumers which translates to revenue share of the entire market for the producers (for that product or service). The tremendous investments made by firms in AI (Tobin and Karaian, 2025; Waters and Bradshaw, 2025) seem to support this grander mission. In this sense, the use of Jevons' Paradox, a statement about price elasticity, could be viewed as a sleight-of-hand. A more indirect way to signal an AI capable of substantial displacement of human labor, a potentially less popular statement. The goal of this paper is to explore the meaning of Jevons' Paradox as industry sources use it given the context of the massive investments that have been made in AI. We do not try to predict their success or failure in their endeavors, but to understand the motivation of this ambition.

For simplicity, we begin with a Constant Elasticity of Substitution (CES) utility function (Arrow et al., 1961) to address the question of Jevons' Paradox in the context of AI competing with human labor, and whether the quantity demanded of an AI-provided good rises relatively more than its price decreases. However, the CES approach holds the elasticity of substitution constant, but a salient aspect of AI is its rapidly evolving capabilities which suggests an elasticity of substitution that rises over time.

Accordingly, we extend our analysis using a Variable Elasticity of Substitution (VES) framework to explore the effects of exponential declines in the marginal cost of effective computation  which, via competition and scaling laws, leads to declines in the price and increases in the quality of AI-provided goods and services.  The combination of exponential declines in the price of AI-provided goods and services and increases in quality (increased ability to substitute)  leads to time-varying AI adoption with an  early phase dominated by price, transitioning to a mature phase based on quality increases to achieve market domination. In fact, unless AI quality rises sufficiently (elasticity of substitution above 1), no level of AI price declines (provided $p_{A}>0$) will result in complete capture of markets.

We explore generalizations which allow for slower power law quality improvements and price declines. In addition, we allow for uncertainty in the quality improvements and heterogeneity in the initial elasticity of substitution. Finally, we show how examining differences in log-odds of market shares can simplify possible analysis.

 This paper makes four contributions: (1) it translates the concept of Jevons' Paradox (own-price elasticity of less than $-1$) into the implied elasticity of substitution associated with the idea of rising revenue share in response to lower prices of AI; (2) derives analytical conditions for market transformation under a time-varying elasticity of substitution (VES) framework motivated by the neural scaling laws; (3) generalizes the VES framework to allow for power law quality improvements and price declines as well as allowing for uncertainty in quality improvements and heterogeneity in elasticities of substitution; and (4) proposes a five-phase adoption taxonomy highlighting when Jevons' Paradox emerges.

 Section~\ref{jevons} establishes a static CES framework to explore Jevons' Paradox, while Section~\ref{VES} introduces a dynamic VES model to capture AI's evolving role with generalizations.  Section~\ref{illustration} uses specific parameters to depict some possible trajectories. The conclusion in Section~\ref{conclusion} ties these insights together and outlines future research directions.

\section{Jevons' Paradox and Elasticity}\label{jevons}

The constant elasticity of substitution (CES) utility function provides a means of studying how the elasticity of substitution affects the trade-off between consumer choices. We begin with the weighted CES utility function for consumers choosing between human-produced (H) and AI-produced (A) goods in \eqref{utility},

\begin{align}
U &= \bigl((1-\alpha)\cdot H^{\rho} +\alpha\cdot A^{\rho}\bigr)^{1/\rho}\label{utility}
\end{align}

\noindent
where $\alpha \in (0,1)$ represents the relative preference weight for AI-produced goods over human-produced goods, $\rho = \frac{\sigma-1}{\sigma}$, and $\sigma$ represents the elasticity of substitution. Note, we later allow the elasticity of substitution to vary over time ($\sigma(t)$) which would lead  to a variable elasticity of substitution utility function or VES.  However, we first wish to develop the simpler CES case.

Given normalized human prices ($p_H = 1$) and AI prices $p_{A} \in (0,1)$, the AI expenditure or revenue share, $r_{A}$, appears in \eqref{rev_share}.\footnote{The ratio of the AI price to human prices is not that easy to quantify. First, AI prices involve the cost of the tokens for input and output. Second, queries differ across numbers of tokens required. Third, many queries also involve a human and so the AI cost is really a hybrid AI-human cost. Fourth, many tasks may require multiple queries. Finally, the cost of a human query also varies by task. Therefore, the ratio of the AI price to the human price is not obvious in magnitude.}

\begin{align}
r_{A}
&= \frac{\alpha\cdot p_{A}^{1-\sigma}}
       {(1-\alpha)+\alpha\cdot p_{A}^{1-\sigma}}
= \frac{1}{1 + \displaystyle\frac{1-\alpha}{\alpha}\cdot p_{A}^{\sigma-1}}
\label{rev_share}
\end{align}

\noindent
We focus on expenditure share from the consumers which translates to revenue share for the producers (for that product or service) as the industry statements as well as tremendous investments seem to aim for market domination.

What happens to the revenue share as the price of AI declines to low levels?
If the elasticity of substitution $\sigma$ exceeds~1, the revenue share
$r_{A}$ approaches~1. If $\sigma=1$, we have $r_{A}=\alpha$ for any AI price.
If $\sigma<1$, then for every strictly positive price with $0<p_{A}<1$
we obtain $r_{A}<\alpha$; \textit{a fortiori}, as $p_{A}\to 0$ we have
$r_{A}\to 0$. This reveals that no matter how low the AI price falls, if AI does not have sufficient quality to substitute for human products, services, or tasks, it cannot achieve market domination in terms of revenue share $r_{A}$.

To aid understanding of the role of price and the elasticity of substitution consider the elasticity of the revenue share $r_{A}$ to the AI price $p_{A}$ in \eqref{rev_elasticity}.

\begin{align}
E_{r_A,p_{A}} &= \frac{\partial\ln r_{A}}{\partial\ln p_{A}} =-(\sigma-1)\cdot (1-r_{A})\label{rev_elasticity}
\end{align}

\noindent
Clearly, the revenue rises as price decreases when $\sigma>1$ and falls when $\sigma<1$. 

Although the CES model  quantifies the revenue market share as a function of price for a given level of elasticity of substitution $\sigma$, it seems restrictive in that if $\sigma\le 1$, $r_{A} \le \alpha$ and low levels of $p_{A}$ cannot break through this level. Insofar as the models have continued to rise in performance, as measured by benchmarks, this suggests that a model needs to allow $\sigma$ to evolve to reflect the greater model capabilities. To address this, we adopt a variable elasticity of substitution (VES) model in Section~\ref{VES}.

\section{Exponential Declines in the AI Price, Scaling Laws, and Increased  Elasticity of Substitution  (VES)}\label{VES}

This analysis examines AI market penetration dynamics under the further assumption of exponential growth in computational capacity and exponential decline in  AI prices. We derive closed-form solutions that demonstrate the interaction between computational growth rates and price dynamics in determining market transitions.

Assume computational capacity grows exponentially as in \eqref{comp_growth},

\begin{align}
\ln\left(\frac{C(t)}{C(0)}\right) &= g \cdot t \label{comp_growth} 
\end{align}

\noindent
where  $g$ is the computational growth rate.  As mentioned earlier, Ho et al. (2024) document that the compute required to reach a given level of performance for large language models declines by 50\% every 8 months. This translates to a gain in effective computational capacity of $1.04$ in terms of continuous compounding per year. Accordingly, one could entertain a prior around $g \approx 1$. This ignores  the substantial gains from better hardware as well as the large amounts of investment in AI. Against this background, the ``scaling laws'' such as Kaplan et al. (2020) and Hoffmann et al. (2022) indicate that AI capabilities rise logarithmically with computation. In these works computation is  proportional to the number of parameters times the data and thus computation implies scaling of these.

Although most economic models rely on constant elasticity of substitution (CES) models to capture trade-offs among choices, a CES approach does not naturally  capture the increasing capabilities of AI systems over time. The use of  a variable elasticity of substitution (VES) framework addresses this (Revankar, 1971).

Specifically,   assume that improvements in AI capabilities increase the  elasticity of substitution by means of the scaling laws which show a linear improvement in AI performance (lower loss) as inputs (in this case computation) increase logarithmically as in \eqref{sigma_linear}. Although one could  propose other functional forms besides the proportional specification in \eqref{sigma_linear}, we select this specification for simplicity and because it may approximate a more elaborate specification. Note, the actual value of $\delta>0$   could differ across goods, services, tasks, contexts, or organizations.

\begin{align}
\sigma(t) &= \delta \cdot \ln\left(\frac{C(t)}{C(0)}\right) \label{sigma_linear}
\end{align}

\noindent
Due to the exponential growth in computation in \eqref{comp_growth}, \eqref{sigma_linear} reduces to  \eqref{sigma_linear2}.

\begin{align}
\sigma(t) & = (\delta \cdot g )\cdot t=\phi \cdot t, \text{ where } \phi=\delta \cdot g \label{sigma_linear2}
\end{align}

\noindent
This yields an elasticity of substitution in \eqref{sigma_linear2} where the constant $\phi=\delta \cdot g$ represents the compound effect of elasticity of substitution gain through greater computation times the rate of additional computation.\footnote{As long as the elasticity of substitution rises with computation (scaling laws) and computation rises over time (Moore's Law for hardware, Erdil and Besiroglu (2022)  for software, Ho et al. (2024) for language models), the exact form of the relation does not affect the outcome, only the timing.  A linear or concave relation seems conservative. However, if we rely on Sam Altman's assertion (point 3) that ``The socioeconomic value of linearly increasing intelligence is super-exponential in nature.'', $\sigma(t)$ could even rise at a high rate. }

The compound effect parameter $\phi$ captures two  forces driving AI adoption. Namely, the rate at which computation increases ($g$) and the sensitivity of the elasticity of substitution $\delta$ to  computational improvements. In other words, $\phi$ specifies the elasticity of substitution growth rate.  This form links the technical AI scaling laws literature with economic substitution theory.  Note, the models improve in terms of loss with respect to inputs at different reported rates. Kaplan et al. (2020) reported a rate of loss decline of around $0.05$. So if this directly translated into $\delta$, one could entertain a prior on $\delta$ of around $0.05$. However, as mentioned above, this likely varies substantially across various uses.\footnote{We use the word ``prior'' in conjunction with the parameters $g$, $\delta$ to convey a sense of uncertainty as to their values. The model treats the parameters as constants. One could certainly place an informative prior on these parameters and simulate across the potential outcomes which might yield a more dispersed, non-symmetric distribution. We do not pursue this further as the goal is create a simple framework while qualitatively acknowledging uncertainty.} Note, at small, strictly positive $t$ the AI and human shares are almost fixed as $\sigma$ is near 0 and this implies almost perfect complements. However, as $t$ rises  $\sigma$ becomes increasingly positive and this artifact vanishes.

The simple form of \eqref{sigma_linear2} means that one can solve for the time when $\sigma=1$, denoted at $t^{*}$ in \eqref{tstar}.

\begin{align}\label{tstar}
t^{*}&=\dfrac{1}{\phi} 
\end{align}

\noindent
Moreover, the time for any particular value of $\sigma$ is linear and so $\sigma=2$ implies $t=2 \cdot t^{*}$ and so forth.

The previous equations dealt with improvements in AI quality leading to greater elasticities of substitution. However, AI pricing affects adoption and producer revenue. Consistent with Ho et al. (2024), assume exponentially declining AI prices  in \eqref{price}, 

\begin{align}
p_{A}(t) &= p_{A}(0) \cdot \exp(-d \cdot t) \label{price}
\end{align}

\noindent
where $0<p_{A}(0) < 1$ represents the initial price ratio between AI and human goods.\footnote{Although we model the declining marginal  cost of computation and the increased quantity of computation separately, they obviously are related through the demand equation. However, there are other shifters of demand and so we will retain the simplicity of separate equations. In addition, to the degree that AI work, especially when it has a low elasticity of substitution for human work, in reality is hybrid AI-human work. Therefore, pecuniary AI price declines may overstate the price declines in actual tasks.}

Substituting this modified price function \eqref{price} into equation \eqref{rev_share} while maintaining the variable elasticity of substitution from equation \eqref{sigma_linear}, leads to a generalized logistic function \eqref{gen_share_mod} that represents the revenue share $r_{A}(t)$  based on  parameter values for $a, b, c$.

\begin{align}
r_A(t)
&= \dfrac{1}{1 + \exp\bigl(a + b\cdot t + c \cdot t^2\bigr)} \label{gen_share_mod}\\
a&=\ln \biggl(\frac{1-\alpha}{\alpha}\biggr)-\ln p_{A}(0),
\ b = d + \phi\cdot \ln p_{A}(0), \ c = -d\cdot \phi \nonumber
\end{align}

\noindent
 Arguably, $c$  represents the most important term in \eqref{gen_share_mod}. First, it captures three of the most important influences on AI competitiveness: $d$ the rate of price declines; $g$ the growth rate in computational capacity; and $\delta$ the growth rate in quality with respect to the rate of computational growth. Since $d, g, \delta >0$ and $c=-(d \cdot g \cdot \delta)$, then $c<0$. Second,  because $c<0$,  $t^{2}$ will dominate the other terms for large $t$ and thus the magnitude of $c$ determines the speed of the later part of the transition. Finally, the  additional quadratic term of $c \cdot t^{2}$ allows acceleration beyond the standard logistic's symmetrical 'S' shape. This acceleration could make the AI transition occur abruptly. 
 
Although  the revenue share function for a given level of $\phi$, $r_A(t|\phi)$, provides direct insight into market penetration, transforming it can reveal underlying linearities and simplify comparisons between different scenarios. Specifically, the logit transformation, defined as $\logit(p) = \ln(p/(1-p))$, proves particularly useful.

\subsection{Dynamics under Power-Law Quality Improvement and Price Declines  with Random $\sigma_{0}$ and $\phi$}
\label{sec:power_law_dynamics_random}

In this sub-section we explore generalizations to the deterministic model developed above. Specifically we (1) allow for possible heterogeneity in the elasticity of substitution or the uncertainty of the quality improvement rate parameter $\phi$; (2) for power law growth in $\sigma(t)$ and in price declines $p_{A}(t)$; and (3) show how differencing the log-odds of revenue share for different tasks reduces the  complexity of the  relations. 

We begin by adopting the power law forms for the elasticity of substitution over time in \eqref{eq:pl1} and in price declines in \eqref{eq:pl2} using the definition $\beta_{0}$ in \eqref{eq:beta0def}.

\begin{align}
\sigma(t) &= \sigma_{0} + \phi \cdot t^{k} \label{eq:pl1}\\
p_{A}(t) &= p_{A}(0) \cdot \exp\left(-d \cdot t^{\xi}\right) \label{eq:pl2}\\
\beta_{0} &= \ln p_{A}(0) \label{eq:beta0def}
\end{align}

\noindent
This captures the idea that quality improvements and price decline rates may diminish over time, assuming $k, \xi \in (0,1)$. Now the revenue share $r_A(t)$ is given by \eqref{eq:ra1} in terms of $X(t)$ defined in \eqref{eq:xi1}, the complement of which equals the logit of the revenue share over time in \eqref{eq:logitdef}.

\begin{align}
r_A(t) &= \left(1 + \exp\left(X(t)\right)\right)^{-1} \label{eq:ra1}\\
X(t) &= \ln\left(\left(1-r_A(t)\right)/r_A(t)\right) \label{eq:xi1}\\
\logit r_A(t)&=-X(t) \label{eq:logitdef}
\end{align}

\noindent
One can rewrite  $X(t)$  in \eqref{eq:affine_X_onephi} as a linear combination of  the  baseline elasticity parameter $\sigma_0$ and the quality improvement parameter $\phi$ using the definitions for $\Theta(t)$ in \eqref{eq:theta1}, $\Upsilon(t)$ in \eqref{eq:upsilon2}, and $\Xi(t)$ in \eqref{eq:xi2}

\begin{align}
X(t) &= \Theta(t) + \Upsilon(t)\cdot\sigma_{0} + \Xi(t)\cdot\phi
\label{eq:affine_X_onephi} \\
\Upsilon(t) &= \beta_{0} - d \cdot t^{\xi} \label{eq:upsilon2}\\
\Theta(t) &= \ln\left(\frac{1-\alpha}{\alpha}\right) -  \Upsilon(t)\label{eq:theta1}\\
\Xi(t) &= t^{k} \cdot \Upsilon(t)
\label{eq:xi2}
\end{align}

\noindent
This provides the apparatus for examining the generalizations to follow.

\subsubsection{Task elasticity of substitution and quality improvement heterogeneity and uncertainty}

Because equation \eqref{eq:affine_X_onephi} is linear in $\sigma_{0}$ and $\phi$, this facilitates letting these  be random variables instead of constant parameters as in \eqref{eq:Xrv}. 

\begin{align}
\tilde X(t) &= \Theta(t) + \Upsilon(t)\cdot\tilde\sigma_{0} + \Xi(t)\cdot\tilde\phi
\label{eq:Xrv} 
\end{align}

\noindent
This allows modeling some task heterogeneity by perhaps assuming a exponential distribution where some tasks are inherently easy to substitute while a great number start off near 0. It also allows for modeling uncertainty in the distribution for $\tilde\phi$. However, regardless of the distribution, the expectation of $X(t)$ in \eqref{eq:mean_X_onephi} is linear in the individual expectations of $\tilde\sigma_{0}$ and $\tilde \phi$.

\begin{align}
\E\left[X(t)\right] &= \Theta(t) + \Upsilon(t) \cdot \E\left[\sigma_{0}\right] + \Xi(t) \cdot \E\left[\phi\right].
\label{eq:mean_X_onephi}
\end{align}

\noindent
If we assume $\phi$ is common across tasks (i.e., non-random or its mean is used) and only $\sigma_0$ is random, \eqref{eq:var_X_onephi} shows the variance of $X(t)$.

\begin{align}
\operatorname{Var}\left[X(t)\right] &= \Upsilon(t)^{2} \cdot \operatorname{Var}[\sigma_{0}].
\label{eq:var_X_onephi}
\end{align}

\subsubsection{Power law quality improvements and price declines}

In this more general framework, the long-run behavior of the AI revenue share $r_A(t)$ depends on the limiting behavior of $X(t)$.  We further expand $X(t)$ in \eqref{eq:Xrv2} and repeat the definition of $\Upsilon$ in \eqref{eq:upsilon2a}. We solve for $\sigma(t)=1$ as shown in \eqref{eq:pltstar} formed by substituting $\sigma_{t}=1$ into \eqref{eq:pl1} and solving for $t_{G}^{*}$,  the generalized timeline. At that point, $(t^{*}_{G})^{k}\cdot \phi=(1-\sigma_{0})$ and so the term in \eqref{eq:Xrv2} associated with $\Upsilon(t)$, $(\sigma_{0}-1+t^{k} \cdot\phi)=0$. For $t>t_{G}^{*}$,  $(\sigma_{0}-1+t^{k} \cdot\phi)>0$. Since $\beta_{0}<0$, $\Upsilon(t)<0$ and becomes more negative with increasing $t$. Although for $\alpha<0.5$, $\ln\left(\frac{1-\alpha}{\alpha}\right)>0$, this stays constant over time. Therefore,  for sufficiently large $t$, $X(t)<0$ and  becomes progressively more negative with $t$, leading to a greater AI market share. 

\begin{align}
\Upsilon(t) &= \beta_{0} - d \cdot t^{\xi} \label{eq:upsilon2a}\\
 X(t) &= \ln\left(\frac{1-\alpha}{\alpha}\right)  + \Upsilon(t)\cdot(\sigma_{0}-1+t^{k} \cdot\phi)
\label{eq:Xrv2} \\
t_{G}^{*}&=\left( \dfrac{1- \sigma_{0}}{\phi} \right)^{1/k} \label{eq:pltstar}
\end{align}

However, the rise in market share could become slow, depending on the values of $k,\phi$ given the nature of power laws. For example, if $k=0.5$, it would cause the timeline to square, all else equal, and so 10 years if $k=1$ becomes 100 years when  $k=0.5$. If quality improvements grow slowly $\phi$ becomes small which also increases the timeline, perhaps dramatically. Note, allowing for $\tilde\sigma_{0}$ and $\tilde\phi$ in \eqref{eq:pltstar} is possible, which would create $\tilde t_{G}^{*}$ which could have a complicated distribution given that it is the ratio of random variables.

\subsubsection{Simplification from differencing revenue share log-odds across tasks}

As a final generalization,  consider two tasks, indexed $j \in \{1,2\}$,  differing only in their baseline elasticities $\sigma_{0,1}$ and $\sigma_{0,2}$ while sharing the same parameters $\alpha, \beta_0, d, k, \xi,$ and $\phi$. Let $X_j(t)$ be the exponent term for task $j$. The difference is appears in \eqref{eq:gap_X_onephi}.

\begin{align}
\Delta_X(t) &= X_1(t) - X_2(t) \notag \\
&= \left(\Theta(t) + \Upsilon(t)\cdot\sigma_{0,1} + \Xi(t)\cdot\phi\right) - \left(\Theta(t) + \Upsilon(t)\cdot\sigma_{0,2} + \Xi(t)\cdot\phi\right) \notag \\
&= \Upsilon(t) \cdot \left(\sigma_{0,1}-\sigma_{0,2}\right).
\label{eq:gap_X_onephi}
\end{align}

\noindent
The relative AI revenue shares of the two tasks are driven by $\Delta_X(t)$. This differences out $\Theta(t)$ and $\Xi(t)\cdot\phi$ which simplifies the analysis.

\clearpage

\subsection{Five Phase Categorization}

\noindent
Based on the model and prior discussion, as an aid to categorization, AI adoption could be thought of as occurring in  five  phases as presented below:

\begin{enumerate}

\item Phase 1 represents the low substitution dominated regime where $\sigma(t) < 1$ since $p_{A}<p_{H}=1$. AI enhances rather than replaces human production, with adoption driven primarily by initial price advantage. Specifically, when $t$ is near 0  market penetration is driven by both price effects and the initial low market share.  In other words, AI firms must try to begin with a low price to initiate or jumpstart the market. During this phase, perhaps a long phase, AI goods, services, or tasks and human counterparts can be jointly used. Even if $p_{A}(t)$ is almost 0, human goods, services, or tasks will have positive share as $r_{A}<\alpha$. Jevons' Paradox does not occur during this phase.  At the moment, AI seems to improve, but not replace, human abilities in tasks such as coding, and this could indicate an $\sigma<1$.\footnote{In the October 29th 2024 earnings call, Google CEO Sundar Pichai stated ``We're also using AI internally to improve our coding processes, which is boosting productivity and efficiency. Today, more than a quarter of all new code at Google is generated by AI, then reviewed and accepted by engineers.'' In Patel (2025), Nadella states his beliefs in human-AI complementarity, but  Nadella (2025) also promotes Jevons' Paradox which indicates they are substitutes. Although both could be true, given factors outside of the simple model here via large-scale economic growth. Altman (2025) often suggests AI will eventually exit this first phase.} The length of this phase matters as it suggests human training to use AI will aid productivity. However, if this is short, training may not be an optimal investment and firms may turn to  selection (hiring AI effective workers and firing ineffective workers) to increase AI-labor productivity.

\item Phase 2 marks the critical transition point where $\sigma(t) = 1$, corresponding to unitary elasticity of substitution. This occurs when $t^{*}=\phi^{-1}$. This marks the tipping point where the elasticity of substitution between AI and human goods, services, or tasks is equal so that $\sigma(t)=1$. This phase exhibits properties analogous to Cobb-Douglas production, with constant expenditure shares across AI and human inputs despite changing relative prices. In fact, this corresponds to $r_{A}=\alpha$, which yet may be a small market share. Note, this occurs regardless of the price trajectory. At this point, Jevons' Paradox is at the cusp of occurring. 

\item Phase 3 initiates true Jevons' dynamics in revenue share  where $\sigma(t) > 1$ happens when $t > t^{*}$.   Jevons' Paradox comes into play and price decreases actually increase revenue by spurring more usage. As $t$ increases, the term $-d \cdot \phi \cdot t^{2}$ dominates, representing the increasing influence of computational scaling and lower prices on market transformation. In this phase the AI market share is between $\alpha$ and 1.

\item Phase 4 begins when $\sigma(t) \ge 2$ at  $t \ge 2 \cdot t^{*}$ and thus accentuates Jevons' dynamics in revenue share. In the beginning of this phase, when $\sigma=2$, the revenue share shows a simple relation in \eqref{s1} with $p_{A}$, the price of AI, based on substitution of $\sigma=2$ into \eqref{rev_share}. Note, if $p_{A}$ is already low by the time $\sigma=2$, the revenue share $r_{A}$ may have already risen close to 1. The converse of \eqref{s1} appears in \eqref{s2}  where the price $p_{A}$ shows a simple relation with revenue share, $r_{A}$. Also, the elasticity of revenue share to price $E_{ r_{A}, p_{A} }$ has a simple form in \eqref{s3}, based on substitution of $\sigma=2$ into \eqref{rev_elasticity}. For low values of $p_{A}$, the elasticity may  already have low magnitude as lowering the price more will not bring forth more revenue.

\begin{align}
r_{A} 
&= \frac{\alpha}{(1-\alpha)p_{A} + \alpha}
\label{s1}\\
p_{A}&= \left(\frac{\alpha}{1-\alpha} \right) \cdot \left (\frac{1-r_{A}}{r_{A}}\right )\label{s2}\\
E_{r_A,p_A}
&=\frac{\partial\ln r_{A}}{\partial\ln p_{A}}
=-\bigl(\sigma -1\bigr)(1 - r_{A})
\Big|_{\sigma=2}
\;=\;
-\bigl(1 - r_{A}\bigr)
\label{s3}
\end{align}

Of course, as time progresses and the elasticity of substitution $\sigma(t)$ increases beyond 2 while the price $p_{A}$ continues to decrease, the revenue share will continue its ascent. 

\item Phase 5 represents market saturation within the constrained partial equilibrium framework, where $r_{A}=1$. At this boundary, Jevons' Paradox (as applied in terms of revenue share) no longer applies as the elasticity of revenue share goes to 0.  Of course, if AI spurs economic growth the overall market could expand and lead to actual revenue to continue to rise. This sentiment has been put forth by industry leaders, but addressing this lies outside the scope of this  simple model.

\end{enumerate}

\clearpage
\section{Illustration}\label{illustration}

\begin{figure}[htbp]
  \centering
  \textbf{AI Revenue Share Dynamics with Time-Varying Elasticity of Substitution}
  \includegraphics[width=\textwidth]{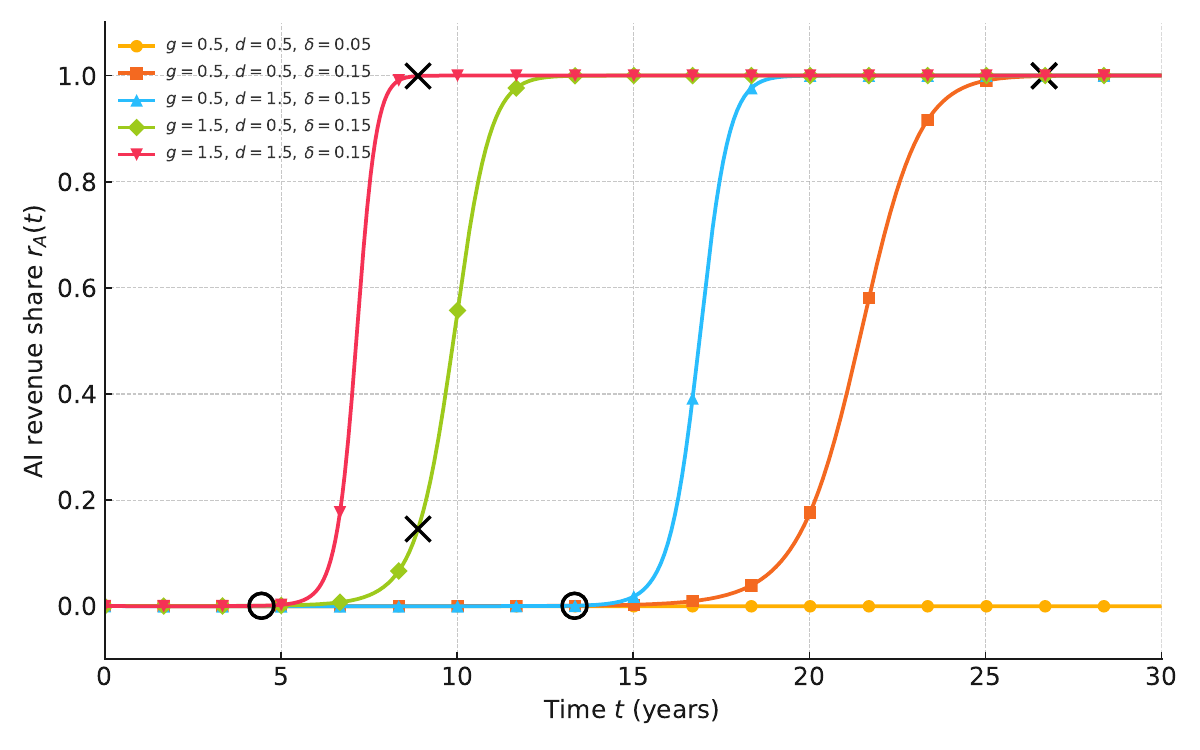}
  \caption[AI revenue share under selected parameter combinations]%
  {Revenue-share paths generated from \eqref{gen_share_mod} with $\alpha = 0.001$ and $p_{A}(0)=0.5$.  After pruning visually indistinguishable cases, the surviving parameter sets are $(g,d,\delta)\in\{(0.5,0.5,0.05),\,(0.5,0.5,0.15),\,(0.5,1.5,0.15),\,(1.5,0.5,0.15),\,(1.5,1.5,0.15)\}$.  Hollow markers depend only on the elasticity path $\sigma(t)=\phi \cdot t$ with $\phi=\delta \cdot g$: a circle at $t^{\ast}=1/\phi$ marks $\sigma=1$ (entry into the Jevons regime), while a cross at $2 \cdot t^{\ast}$ marks $\sigma=2$ (strong Jevons).  The price-decay rate $d$ shifts entire curves without moving these phase boundaries, separating the quality-driven phase structure from the price-driven climb in $r_{A}(t)$.}
  \label{fig:rA_paths}
\end{figure}

To make the preceding analysis more immediate, we provide Figure~\ref{fig:rA_paths} which shows the trajectory of various levels of AI revenue share as a function of the initial preference weight $\alpha=0.001$, initial price $p_{A}(0)=0.5$, and five visually distinct combinations of computation growth parameter $g$, price decline parameter $d$, and AI learning rate $\delta$ over a 30 year period. We selected the initial preference weight $\alpha=0.001$ for two reasons. First, this allows showing a transition or adoption curve going over the full range from $r_{A}(t)$ close to 0 for low $t$ and close to 1 (in some cases) for large $t$. Second, AI has always had a low pecuniary price, even at low adoption rates. The unusual nature of AI means that for an equal pecuniary price, it would have low rates of adoption. This could change over time with familiarity.

\clearpage

Before the circle marker the system is in Phase~1, where AI and human inputs are complements and Jevons' Paradox does not arise.  Between the circle and the cross ($1<\sigma<2$) the system enters Phase~3, exhibiting elastic demand and Jevons dynamics.  To the right of the cross ($\sigma>2$) it moves into Phase~4, a ``strong'' Jevons regime.  Phase~2 is a  knife-edge at $\sigma=1$ and Phase~5 is the limiting saturation as $r_{A}(t)\to1$.  Thus the figure illustrates how identical substitution paths can yield different revenue trajectories when price declines occur at different speeds.

 The two black markers depend only on the compound rate $\phi=\delta \cdot g$.  Doubling $g$ (faster compute growth) or doubling $\delta$ (a steeper scaling law) halves the time needed to reach each phase boundary.  Hence curves with the same $\phi$ share the same phase times regardless of how the product is split between the two components.  By contrast, the price-decay rate $d$ merely translates the entire $r_{A}(t)$ path left or right without changing $t^{\ast}$ or $2 \cdot t^{\ast}$.  When $\phi$ is large even a modest $d$ yields a near-vertical takeover; when $\phi$ is small, aggressive price cuts (large $d$) are required just to move off the horizontal axis.  In short, $\phi$ sets the clock for substitution, whereas $d$ determines how quickly revenue follows once substitution becomes technologically feasible.

The figure thus separates the
\emph{quality-driven phase structure} (dictated by $\sigma(t)$) from the \emph{price-driven} rise in $r_{A}(t)$, illustrating how rapid cost declines ($d$ large) can accelerate revenue capture even when the elasticity path $\sigma(t)$ is common across scenarios.

Because such interest exists on AI timelines, we examine all the parameter combinations in Table~\ref{tab:rA_summary}.  It displays the compound rate $\phi=\delta \cdot g$, the phase~2  boundary $t^{\ast}=1/\phi$, and the AI revenue share at $t=10$ and $t=30$~years.  It reveals cases where AI never dominates (rows 1,3 ) and those where it does in as little as 10 years (last row).

\begin{table}[htbp]
\centering
\sisetup{table-number-alignment = center}
\begin{tabular}{
    S[table-format=1.1]  % g
    S[table-format=1.1]  % d
    S[table-format=1.2]  % delta
    S[table-format=1.3]  % phi
    S[table-format=2.3]  % t*
    S[table-format=1.6]  % rA(10)
    S[table-format=1.6]  % rA(30)
}
\toprule
{$g$} & {$d$} & {$\delta$} & {$\phi$} & {$t^{\ast}$} &
{$r_{A}(10)$} & {$r_{A}(30)$}\\
\midrule
0.5 & 0.5 & 0.05 & 0.025 & 40.000 & 0.000014 & 0.000020\\
0.5 & 0.5 & 0.15 & 0.075 & 13.333 & 0.000241 & 0.999997\\
0.5 & 1.5 & 0.05 & 0.025 & 40.000 & 0.000000 & 0.000000\\
0.5 & 1.5 & 0.15 & 0.075 & 13.333 & 0.000020 & 1.000000\\
1.5 & 0.5 & 0.05 & 0.075 & 13.333 & 0.000241 & 0.999997\\
1.5 & 0.5 & 0.15 & 0.225 & 4.444 & 0.552229 & 1.000000\\
1.5 & 1.5 & 0.05 & 0.075 & 13.333 & 0.000020 & 1.000000\\
1.5 & 1.5 & 0.15 & 0.225 & 4.444 & 0.999997 & 1.000000\\
\bottomrule
\end{tabular}
\caption{Phase-transition time $t^{\ast}$ and AI revenue shares for each
parameter triple, computed with $\alpha = 0.001$ and $p_{A}(0)=0.5$.}
\label{tab:rA_summary}
\end{table}

We present some of the key insights below:

\begin{enumerate}
\item \textbf{Phase timing depends on $\phi=\delta \cdot g$.}
      Rows with the same $\phi$ share the same $t^{\ast}$, regardless of
      the price-decay rate~$d$.
\item \textbf{Price cuts affect revenue primarily after $\sigma>1$.}
      Compare rows with identical $\phi$ but different $d$, such as row two and row four. The  identical $\phi=0.075$ gives the same
      $t^{\ast}$. However, the effectiveness of aggressive price cuts (a higher $d$) in accelerating revenue share growth is primarily realized after $\sigma>1$. Before this threshold (i.e., when $t<t^{*}$), a higher $d$ can, under certain conditions illustrated here, lead to a temporarily lower revenue share $r_{A}(t)$ compared to a scenario with slower price declines.

\item \textbf{Large $\phi$ ensures rapid saturation.}
      When $\phi=0.225$ the revenue share already exceeds
      $55\%$ by year 10 and essentially saturates by year 30,
      independent of~$d$.
\item \textbf{Small $\phi$ plus aggressive price cuts is ineffective.}
      The combination $\phi=0.025$, $d=1.5$ leaves
      $r_{A}(30)\approx 0$; Jevons' Paradox never materializes without
      faster compute growth or a steeper scaling law.
\end{enumerate}

Examination of the Figure and Table reveal that  investment that raises either the compute growth rate $g$ or the
scaling-law sensitivity $\delta$ offers far greater leverage over AI
market penetration than slashing prices~$d$ alone.  Price competition matters chiefly after technology makes AI and human inputs substitutable. In other words, AI quality is the most important factor in the AI transition.

\clearpage
\section{Conclusion}\label{conclusion}

Frequent invocation of Jevons' Paradox by AI leaders may appear as typical industry  jargon   devoid of much meaning, but as shown here it equates to stating that AI can substantially or completely substitute for human provision of goods, services, or tasks. Given the  importance of  Jevons' Paradox to the progress of AI, it seems timely to examine it in more detail.

 This analysis provides a framework that incorporates these statements  to make the underlying logic clearer. Specifically, we examine market transitions to AI-produced goods using an elasticity of substitution framework. We introduce a variable elasticity of substitution (VES) model where the increased performance of AI based on the scaling laws expands as a function of computational growth that leads to the elasticity of substitution $\sigma(t)$ growing over time. The growth in $\sigma(t)$ coupled with falling prices for AI driven by competition and the lower marginal costs of computation  (stemming from hardware, algorithmic, and software progress) create an environment where AI becomes better and more inexpensive relative to humans over time. This can still remain true under less restrictive assumptions, but these may lengthen the timelines.
 
As a categorization of  AI progress, we set forth a five-phase model where (1) initial adoption depends critically on sufficiently low starting prices relative to substitution sensitivity; (2) transition points between phases are determined by parameters linking computational scaling to substitution elasticity; (3) Jevons' Paradox occurs in terms of increases in revenue with price decreases when $\sigma(t) > 1$; (4) A stronger form of Jevons' Paradox occurs  when $\sigma(t) > 2$; and (5) market saturation eventually leads to unitary price elasticity (no Jevons' Paradox).

In Section~\ref{illustration} we provided some specific scenarios to make the analysis more immediate and illustrated these via  Figure~\ref{fig:rA_paths} and Table~\ref{tab:rA_summary}. These scenarios underscore a simple hierarchy.  First, the compound rate $\phi=\delta \cdot g$ dictates when
the economy reaches the substitution thresholds $\sigma=1$ and
$\sigma=2$; doubling either the compute-growth rate $g$ or the scaling-law coefficient $\delta$ halves the time to each phase boundary.  Second, once $\sigma>1$, the price decay rate $d$ controls how fast revenue flows to AI providers along an otherwise fixed substitution path. Finally, no amount of price cutting can compensate for a low $\phi$: when $\delta \cdot g$ is too small the Jevons regime may not arrive within any relevant planning horizon.  For policymakers and investors this means that subsidizing algorithmic or hardware efficiency is a more potent lever than temporary price support, while for firms it suggests that competition on quality dominates price competition until AI and human inputs become \textit{bona fide} substitutes. 

However, since the translation of the gain from extra computation has a different effect by task, product, or service, the curves or scenarios depicted may also represent different share trajectories across individual tasks, products, or services. Insofar, as some scenarios never show an AI takeoff (perhaps martial art matches) while others show quick domination (perhaps coding), this indicates some of the disruption associated with AI will stem from optimization across  continual shifting prices and capabilities. 

For any given scenario, the steeper `S'-curve relative to a standard logistic, driven by the product $c$ of rates of price declines, computational growth, and quality improvements could further exacerbate the disruption making it seem abrupt or unanticipated.

Having presented this model to  understand the perspective of AI proponents, many alternative perspectives exist on the long term growth of AI (Erdil and Besiroglu, 2024). The simple partial equilibrium model presented above does not account for the many positive or negative feedback loops in a more general equilibrium that could accelerate or impede adoption. For example, we held the price of the human product at 1, regardless of the context, whereas this may change in response to AI competition. We also provide a perhaps false choice as AI versus human when hybridization has obvious potential.  We also do not discuss the differential changes in substitution and associated feedback loops that could occur across products, services, or tasks as AI, through the mechanism of high-quality data availability (e.g., programming, science). Similarly, we neglect network and lock-in effects which could vary somewhat the trajectory. We treat quality gains as shifts in $\sigma(t)$ and cost declines as $p_{A}(t)$. Yet scaling laws plausibly do both together (computational gains drive both). In addition, the optimistic view of AI stresses the increase in overall wealth and quality-of-life potential which could feedback into choices. If AI grows income and wealth, usage could continue to grow despite having captured the entire market which could lead to a continuation of  Jevons' Paradox.  In addition, we do not discuss possible impediments to AI development such as declining quantity and quality of data, energy usage,  capital costs, scaling laws failing to hold, frictions, regulation, and other possible bottlenecks. Early progress could be misleading as AI can initially substitute in easy settings, but may encounter a large range of more difficult settings later.

The education sector offers an interesting  counterpoint to tech industry invocations of Jevons' Paradox. A number of public and accredited open universities  exist across the world that  allow credit by examination and other alternative ways of measuring knowledge acquisition  such as Thomas Edison State University (US), the Open University (UK), Indira Gandhi National Open University (IN), and others. These institutions have delivered education at fraction of traditional higher education costs for decades. The internet reduced content distribution costs to essentially zero, yet market transformation never materialized. TESU captures negligible US market share with under 12,000 students and the even  more successful Open University has captured only a plurality of the UK market. Under the proposed framework, this reveals education as a $\sigma < 1$ market where Jevons' Paradox fails to manifest in revenue terms. AI advocates suggest their technology transcends mere pecuniary cost reduction, offering superior knowledge, tireless availability, and adaptive personalization. However, the current analysis indicates that these quality improvements must push the elasticity of substitution above unity before capturing significant revenue share. The success of AI-driven platforms like Duolingo in specific educational niches suggests this threshold is achievable, but the resilience of traditional universities despite enormous cost disadvantages implies that broad educational transformation requires AI to deliver substitutability improvements  exceeding current capabilities. Undoubtedly, part of the current lack of substitution comes from the different experience of being around other people of the same age in similar circumstances that help create substantial network effects for graduates as well as other non-pecuniary aspects of university life. This current Maginot Line of higher education may yet fall as AI finds ways to circumvent it, much as social media created new forms of connection rather than replicating traditional social experiences.

Despite the barriers that AI adoption could face, investor revealed preference indicates their latent, positive AI forecast. For example, Apple recently announced  \$500B of AI related investments (Tobin and Karaian, 2025), and many of the leading firms have set forth \$300B in planned investments  (Waters and Bradshaw, 2025). Supporting their optimism, OpenAI now has over 400 million weekly active users (Mishkin and Nellis, 2025). 

The influence of Jevons (1866), research on scaling laws (Kaplan et al. 2020; Hoffmann et al., 2022), which at the time was ``academic,'' established concepts of marginal cost as well as competition, and Moore's Law (Moore, 1965) may have motivated these large investments which cumulatively exceed \$1T.  Whether this leads to the greatest investment bubble of all time or an age of ``super abundance'' remains to be seen as the scope of the opportunity may increase the scope for possible error. Given the motivation for these investments often invokes well-established economic ideas and technical research, a  quote from  Keynes (1936) may prove apt.

``The ideas of economists and political philosophers, both when they are right and when they are wrong, are more powerful than is commonly understood. Indeed the world is ruled by little else. Practical men, who believe themselves to be quite exempt from any intellectual influences, are usually the slaves of some defunct economist. Madmen in authority, who hear voices in the air, are distilling their frenzy from some academic scribbler of a few years back. I am sure that the power of vested interests is vastly exaggerated compared with the gradual encroachment of ideas.''

\clearpage
\section*{References}\label{refs}

\mybib Altman, Sam (2025), ``Three Observations,'' \textit{blog.samaltman.com/three-observations}.

%\mybib Arrow, Kenneth J. (1962), ``The Economic Implications of Learning by Doing,'' \textit{The Review of Economic Studies}, Vol. 29, No. 3,  pp. 155--173.

\mybib Arrow, K.J., Chenery, H.B., Minhas, B.S., and Solow, R.M. (1961), ``Capital-Labor Substitution and Economic Efficiency,'' \textit{The Review of Economics and Statistics}, Vol. 43, No. 3, pp. 225--250.

\mybib Erdil, Ege and Tamay Besiroglu (2022). Algorithmic progress in computer vision. \textit{arXiv preprint arXiv:2212.05153}.

\mybib Erdil, Ege and Tamay Besiroglu (2024), ``Explosive Growth from AI Automation: A Review of the Arguments,'' arXiv:2309.11690v3.

%\mybib G\o rtz, Erik (1977), ``An Identity between Price Elasticities and the Elasticity of Substitution of the Utility Function,'' \textit{The Scandinavian Journal of Economics}, Vol. 79, No. 4, pp. 497--499.

%\mybib Hicks, J.R. and R.G.D. Allen (1934), ``A Reconsideration of the Theory of Value,'' \textit{Econometrica}, Vol 1., pp. 196--219.

\mybib Hoffmann, Jordan,  Sebastian Borgeaud, Arthur Mensch, Elena Buchatskaya, Trevor Cai, Eliza Rutherford, Diego de Las Casas, Lisa Anne Hendricks, Johannes Welbl, Aidan Clark, Tom Hennigan, Eric Noland, Katie Millican, George van den Driessche, Bogdan Damoc, Aurelia Guy, Simon Osindero, Karen Simonyan, Erich Elsen, Jack W. Rae, Oriol Vinyals, and Laurent Sifre (2022), ``Training Compute-Optimal Large Language Models,'' arXiv: 2203.15556.

%Kaplan paper has 1111 cites
%\bibitem{Kaplan2020}
\mybib Kaplan, Jared, Sam McCandlish, Tom Henighan, Tom B. Brown, Benjamin Chess, Rewon Child, Scott Gray, Alec Radford, Jeffrey Wu, and Dario Amodei (2020), ``Scaling Laws for Neural Language Models.'' arXiv:2001.08361.

\mybib Keynes, John Maynard (1936), \textit{The General Theory of Employment, Interest and Money}, Ch 24, London: Macmillan.

\mybib Mishkin and Nellis (2025), ``OpenAI's ChatGPT exceeds 400 million weekly active users,'' \textit{Reuters}, February 20.

\mybib Moore, G. E. (1965). Cramming more components onto integrated circuits. \textit{Electronics}, 38(8), 114--117.

\mybib Nadella, Satya (2025), ``Jevons paradox strikes again!,'' \textit{X}, Jan. 26.

\mybib Patel, Dwarkesh (2025), ``Satya Nadella -- Microsoft's AGI Plan and Quantum Breakthrough,'' \textit{Dwarkesh Patel Podcast}, February 19.

\mybib Pichai, Sundar. (2024). ``Google Q3 2024 earnings call: CEO Remarks,'' Google  Quarterly Earnings Conference Call, October 24.

\mybib Revankar, Nagesh S. (1971), ``A Class of Variable Elasticity of Substitution Production Functions,'' \textit{Econometrica}, Vol. 39, No.  1, pp. 61--71.

\mybib Rosalsky, Greg (2025), ``Why the AI world is suddenly obsessed with a 160-year-old economics paradox,'' \textit{National Public Radio}, Feb. 4.

\mybib Smith, Talmon Joseph (2025), ``DeepSeek Doesn't Scare OpenAI, Thanks to the 'Jevons Paradox,'' \textit{New York Times}, Feb. 19.

%\mybib Smith, V. Kerry (1979), ``On Relations Between Price Elasticities and the Elasticity of Substitution in Consumption: A Comment,'' \textit{The Scandinavian Journal of Economics},  Vol. 81, Issue 1, pp. 115.

%\mybib Applegate, D. L., Bixby, R. E., Chvátal, V., and Cook, W. J. (2006). \textit{The Traveling Salesman Problem: A Computational Study}. Princeton University Press.

%\mybib Hernandez, D., and Brown, T. B. (2020). Measuring the algorithmic efficiency of neural networks. \textit{arXiv preprint arXiv:2005.04305}.

%\mybib Sutton, R. S. (2019). The bitter lesson. \textit{Incomplete Ideas (blog)}, 13. 

%\mybib Sutter, H. (2011). The free lunch is over: A fundamental turn toward concurrency in software. \textit{Dr. Dobb's Journal}, 30(3), 202-210.

%\mybib Thompson, N. C., Greenewald, K., Lee, K., and Manso, G. F. (2020). The computational limits of deep learning. \textit{arXiv preprint arXiv:2007.05558}.

%\mybib Grace, K. (2013). Algorithmic progress in six domains. Technical report, Machine Intelligence Research Institute.

\mybib Ho, Anson, Tamay Besiroglu, Ege Erdil, David Owen, Robi Rahman, Zifan Carl Guo, David Atkinson, Neil Thompson, Jaime Sevilla (2024), ``Algorithmic progress in language models,'' ArXiv:2403.05812v1.

%\mybib Hobbhahn, M., and Besiroglu, T. (2022). Trends in GPU price-performance. Technical report, Epoch AI.

\mybib Jevons, William Stanley (1866), \textit{The Coal Question}, 2nd ed., London: Macmillan and Company. 

%\mybib Leiserson, C. E., Thompson, N. C., Emer, J. S., Kuszmaul, B. C., Lampson, B. W., Sanchez, D., and  Schardl, T. B. (2020). There's plenty of room at the top: What will drive computer performance after Moore's law? \textit{Science}, 368(6495).

\mybib Tobin and Karaian (2025), ``Apple to Invest \$500 Billion in U.S. as Trump Tariffs Loom,'' \textit{New York Times}, February 24.

\mybib Waters and Bradshaw (2025), ``Big Tech lines up over \$300bn in AI spending for 2025,'' \textit{Financial Times}, February 6.

\end{document}